\documentclass{article}
\usepackage{arxiv}
\usepackage[round,authoryear]{natbib}
\usepackage[utf8]{inputenc} 
\usepackage[T1]{fontenc}    
\usepackage{hyperref}       
\usepackage{url}            
\usepackage{booktabs}       
\usepackage{amsfonts}       
\usepackage{nicefrac}       
\usepackage{microtype}      
\usepackage{lipsum}		
\usepackage{graphicx}
\usepackage{doi}
\usepackage{tabularx}
\usepackage{graphicx}
\usepackage{subcaption}
\usepackage{amsmath}
\usepackage{mdframed}
\usepackage{longtable}
\newcolumntype{P}[1]{>{\raggedright\arraybackslash}p{#1}}
\usepackage{comment}
\usepackage{csquotes}
\usepackage[table]{xcolor}
\usepackage{multirow}
\usepackage{array}
\usepackage{float}

\usepackage{orcidlink}
\usepackage{authblk} 

\usepackage{listings}
\usepackage{soul}
\usepackage{xcolor}

\usepackage{fancyvrb}

\graphicspath{{images/}}

\title{Stabilising Learner Trajectories: A Doubly Robust Evaluation of AI-Guided Student Support using Activity Theory}

\author[1]{Teo Susnjak\orcidlink{0000-0001-9416-1435}}
\author[2]{Khalid Bakhshov\orcidlink{0000-0002-4379-4547}}
\author[1]{Anuradha Mathrani\orcidlink{0000-0002-9124-2536}}
\affil[1]{School of Mathematical and Computational Sciences, Massey University, Auckland, New Zealand}
\affil[2]{Senior Advisor, Student Achievement - Student Services, Massey University, Auckland, New Zealand}

\renewcommand{\shorttitle}{Evaluation of AI-Guided Student Support using Activity Theory}

\begin{document}
\maketitle

\begin{abstract}
While predictive models are increasingly common in higher education, causal evidence regarding the interventions they trigger remains rare. This study evaluates an AI-guided student support system at a large university using doubly robust propensity score matching. 
We advance the methodology for learning analytics evaluation by leveraging time-aligned, dynamic AI \textit{probability of success} scores to match 1,859 treated students to controls, thereby mitigating the selection and immortal time biases often overlooked in observational studies.
Results indicate that the intervention effectively stabilised precarious trajectories, and compared to the control group, supported students significantly reduced their course failure rates and achieved higher cumulative grades. However, effects on the speed of qualification completion were positive but statistically constrained.
We interpreted these findings through Activity Theory, framing the intervention as a socio-technical ``brake” that interrupts and slows the accumulation of academic failure among at-risk students.
The student support–AI configuration successfully resolved the primary contradiction of immediate academic risk, but secondary contradictions within institutional structures limited the acceleration of degree completion. We conclude that while AI-enabled support effectively arrests decline, translating this stability into faster progression requires aligning intervention strategies with broader institutional governance.
\end{abstract}

\keywords{Learning Analytics  \and  Causal Inference \and Early Warning Systems  \and  Propensity Score Matching \and Activity Theory  \and Proactive student support \and Student Retention}

\section*{Highlights}
\begin{itemize}
    \item AI-guided proactive support reduces course failures among students identified as high risk.
    \item Dynamic AI-generated risk scores enable matching procedures that address immortal time bias.
    \item Doubly robust propensity score models indicate improvements in cumulative grades and pass rates.
    \item Activity Theory frames AI-supported interventions as systemic stabilisers of student trajectories
    \item Results provide guidance for revising support rules, operational workflows and governance.
\end{itemize}

\section{Introduction}

With the emergence of AI-driven analytical methods, computational researchers are increasingly turning to learning analytics (LA) to study student engagement and identify gaps that place pressure on retention and success. Recent syntheses and large-scale studies report mixed but promising effects of LA interventions and highlight the centrality of instructional context and student characteristics in shaping impact \citep{Sonderlund2019bjet,Jovanovic2021compedu,Gasevic2016ihe}. LA interventions typically apply statistical and machine-learning techniques to educational data such as Learning Management Systems (LMS) interaction logs, demographics, and prior grades to construct predictive indicators of engagement and risk, and while such predictors support forward-looking instructional design, they often lack explicit theoretical grounding, which limits interpretability and the ability to translate model outputs into sustainable pedagogical practice \citep{Rose2019bjet}. Because the instructional environment is shaped by how institutional tools are contextualised to learner needs in specific settings, effective LA requires acknowledging the dialectic between instruction and learning \citep{Engestrom1987Learning} and designing analytics that take into account both institutional conditions and individual attributes. In parallel, LA work benefits from an organising framework that makes explicit the stakeholders served, objectives pursued, data and instruments used, and the operational constraints. A widely cited framework articulates six such dimensions for LA services \citep{drachsler2012pulse}, and this perspective complements activity-theoretical views of institutional practice \citep{blackler2000organizing}.

Despite rapid adoption of predictive risk models in higher education, there is still limited causal evidence on the effectiveness of AI-guided support when deployed at scale or, especially, when used as \emph{proactive} decision-support systems that enable interventions before major negative events surface \citep{herodotou2020can}. Moreover, even less work has been conducted that connects such evaluations to a theoretically grounded account of how student support interventions are embedded in institutional activity systems \citep{schwendimann2016perceiving}. Many studies focus on predictive accuracy or descriptive outcomes of LA tools while giving less attention to the counterfactual question of what would have happened to at-risk students in the absence of targeted support, or to the organisational arrangements that mediate how predictions translate into action \citep{Rose2019bjet,susnjak2024beyond}. In particular, evaluations are needed that simultaneously (i) approximate the causal effects of LA-triggered interventions under transparent identification assumptions \citep{weidlich2022causal} and (ii) situate those interventions within a socio-technical account of the roles, rules, tools, and objectives that govern the institutional response \citep{Greller2012FrameworkLA}.

This study attempts to address this gap by analysing an AI-based LA intervention for undergraduate students in a large tertiary institution who are identified as \textit{at-risk} of qualification non-completion.
One of the primary motivations for the deployment of predictive analytics in this institution was to shift student support from a predominantly reactive referral-based model to proactive outreach. Traditional services are often activated only when students or teaching staff seek help, typically in response to acute difficulties, and tend to prioritise short-term remediation over prevention \citep{kuh2006matters}. Prior work on student success shows that preventative and proactive approaches are more effective for retention and achievement because they reduce structural and psychological barriers and strengthen students’ sense of belonging \citep{tinto2012completing,drake2011role}. In this context, predictive analytics functions as an enabling mechanism, whereby, risk scores allow the student support team to anticipate emerging difficulties and intervene before they escalate, systematically scaffolding key transition points with targeted resources and opportunities to build self-efficacy and resilience \citep{glennen1985reduction,varney2013proactive}.The institution under study, deployed a machine-learning system \citep{susnjak2024beyond} that generates daily probability of qualification completion/success for each student, and the student support team uses this risk score as one of the main inputs when deciding whom to contact and support. We then built on expansive learning within Activity Theory (AT) \citep{Engestrom1987Learning,Engestrom1999pat}, to use it to analyse the institutional configuration of this AI-enabled intervention alongside causal analysis of its academic effects. Prior LA work in higher education often prioritises prediction while remaining silent on the pedagogical and organisational representation underpinning intervention design \citep{Rose2019bjet}, whereas, this study integrates technical and pedagogical assessment to provide a more comprehensive evaluation of LA intervention practice.

Methodologically, we used the propensity score matching (PSM) \citep{rosenbaum1983central} approach to approximate the counterfactual outcomes of students who received AI-guided support. Propensity scores were estimated from rich pre-intervention covariates, including demographic, enrolment, and prior performance information, as well as the AI-generated probability of success at a pre-specified time point. We then implemented a blocked and calipered matching design, followed by doubly robust outcome estimation with bootstrap uncertainty quantification and sensitivity analysis to unobserved confounding. This design allowed us to estimate how much the intervention changed patterns of failure, pass rates, grades, and qualification progression among students assessed by the institutional AI models as being at elevated risk of non-completion.

AT served as a complementary lens that situated the predictive AI model, central student support unit, policies, and students within a single activity system. AT framed the LA infrastructure and the support team as mediating tools and communities that act on the shared object of improving student success under defined rules and divisions of labour \citep{Engestrom2015Learning,blackler2000organizing}. By mapping our empirical findings onto this framework, we examined how AI-driven analytics, human judgment, and institutional constraints interact, and we reflected on the tensions and opportunities that arise when LA interventions are institutionalised.

The study therefore, had two primary objectives. First, it applied PSM and doubly robust estimation to quantify the association between the AI-guided support intervention and student academic outcomes under clearly stated identification assumptions. Second, it used AT to provide a pedagogical and organisational reflection on the design, implementation, and institutional embedding of the intervention, linking observed effects to the structure and evolution of the underlying activity system \citep{Engestrom2015Learning}.
 Based on these objectives, the study addresses the following research questions.

\begin{description}
  \item[RQ1:] To what extent is the AI-guided support intervention associated with changes in academic performance and qualification progression for at-risk undergraduate students, relative to comparable non-supported peers?
  \item[RQ2:] How does the Activity Theory framework characterise the roles, tools, rules, and divisions of labour involved in the AI-driven LA intervention, and how do these configurations help to explain the observed patterns of impact?
\end{description}

\section{Literature Review}
\label{sec:lit_review}

\subsection{Activity Theory}

\begin{figure}[htbp]
\centering
  \includegraphics[width=\linewidth]{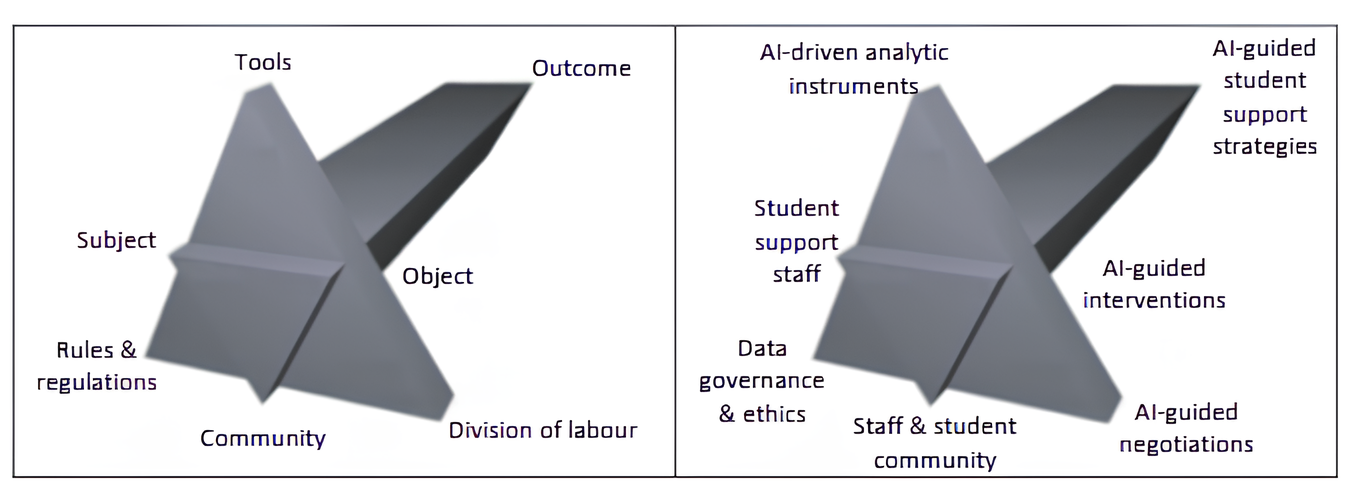}

\caption{The Activity Theory framework(left) applied to the study's learning analytics context (right).}
\label{fig:at}
\end{figure}

AT provides an interdisciplinary framework for studying human activity as a socially mediated, tool-embedded pursuit of shared objects within networks of collaborating entities, and has been used to analyse organisational work and learning \citep{blackler2000organizing,Korpela2002cscw}. Engestr\"om argues that work activities evolve in real situations where actions are contextualised by histories and roles; using mediating tools and technologies, participants reconfigure their circumstances to seek improved outcomes \citep{Engestrom1987Learning,Engestrom1999pat}. In AT terms, an activity comprises actions distributed by a division of labour among subjects who use tools to coordinate within communities governed by rules while working on a shared object that is transformed into an outcome \citep{Engestrom2015Learning}. Figure~\ref{fig:at} (left) depicts the AT framework widely adopted in analyses of psychological, educational, and organisational motives for collective outcomes, and as such, supports reasoning about complex socio-technical systems while identifying collaborating entities across organisational levels \citep{blackler2000organizing,Engestrom1999pat}. Visualising this structure can help surface tensions and inform policies to address them, and it focuses evaluation on mediating mechanisms and bottlenecks that condition outcomes; in educational intervention design, this supports analysis of dyads such as subject–tools and community–division of labour in order to leverage collaboration for improved learning outcomes.

In the context of predictive analytics, adopting an AT perspective shifts attention from purely labelling individual students as ``at risk”, toward diagnosing breakdowns and contradictions in the broader learning system \citep{Engestrom1987Learning,engestrom2001expansive,vygotsky1978mind}. Rather than locating problems within learners, AT interprets difficulties as tensions between subjects, tools, rules, community, and division of labour, such as misalignments between pedagogical practices, assessment structures, and students’ cultural or material conditions \citep{Engestrom1987Learning,engestrom2001expansive}. This systems-oriented view resonates with student-success frameworks that emphasise institutional responsibility for enabling academic and social integration and designing environments that promote high-impact engagement practices \citep{tinto2012completing,kuh2001assessing,kuh2008high,kuh2009national}, and, in this study's context, provides a conceptual bridge between predictive risk scores and the redesign of mediating conditions for learning.

\paragraph{Pedagogical reflection using Activity Theory}

AI-driven analytics uses statistical and machine-learning methods to produce LA applications for monitoring performance and coordinating institutional responses, whereby a generic LA framework specifies six design dimensions: stakeholders, objectives, data, instruments, internal and external constraints—that help structure such work \citep{Greller2012FrameworkLA}. These dimensions inform design and implementation when paired with AT as the theoretical foundation; effects relate to changes in the learning environment and to the development of robust institutional strategies for ongoing support, stakeholders map to the communities of practice that enact semiotic interpretations in defined divisions of labour, rules articulate data governance and ethics, objectives cover monitoring and targeted support, and tools encompass AI-driven analytic instruments such as propensity score methods used alongside other modelling techniques \citep{Engestrom2015Learning,Greller2012FrameworkLA}. Within this structure, student support teams act as subjects who, informed by LA feedback, interact with identified cohorts, facilitate awareness, and sustain engagement, while intervention efficacy is assessed at the individual level and aggregated to guide incremental improvement and institutionalisation of support strategies \citep{Sonderlund2019bjet,Jovanovic2021compedu,Gasevic2016ihe}.
The application and contextualisation of AT to this study are depicted in Figure \ref{fig:at} (right).

\subsection{Artificial intelligence for identifying at‐risk students}

Over the past decade, institutions have increasingly deployed machine‐learning models to flag students at elevated risk of withdrawal or failure with an overwhelming focus on course-level, rather than programme/qualification-level predictions \citep{namoun2020predicting}. These systems typically draw on LMS activity traces, enrolment records, assessment histories, and demographic information as inputs \citep{Sonderlund2019bjet,Jovanovic2021compedu,Gasevic2016ihe}. Such early‐warning systems also tend to use off-the-shelf supervised learning algorithms such as regularised logistic regression, gradient boosting, and random forests to estimate individualised probabilities of success or continuation, which can then inform student support teams and teaching staff through dashboards and alerts. Recent reviews emphasise that the predictive performance of these systems is sufficiently adequate for triage \citep{namoun2020predicting}, but that their impact ultimately hinges on how risk scores are integrated into institutional workflows and human decision‐making \citep{Gasevic2016ihe,Rose2019bjet}. 

\paragraph{Interventions in Higher Education}

Higher education institutions increasingly deploy targeted support interventions to address concerns about progression, retention, and equity, often guided by early–warning analytics that flag students at elevated risk of failure or withdrawal \citep{wong2020review}. While early syntheses reported mixed effects depending on how risk information is raised to student support teams \citep{Sonderlund2019bjet}, recent empirical work has begun to isolate the impact of specific proactive strategies. For instance, \citet{hall2021effects} demonstrated that proactive student-success coaching, informed by predictive analytics, positively influenced persistence in community colleges, though the effect size varied by student demographic. Similarly, \citet{han2025instructor} found that instructor-initiated academic alerts were associated with lower withdrawal rates and higher grades, highlighting the critical role of the human agent in the loop. However, the translation of predictive models into effective action remains a challenge. \citet{alalawi2025evaluating} argued for an ``action framework" where the value of LA is measured not by model \textit{accuracy}, but by the \textit{efficacy} of the subsequent intervention. Despite this, large-scale randomised controlled trials (RCTs) remain uncommon due to ethical and logistical constraints \citep{herodotou2020can}. \citet{herodotou2020can} evaluated a rare large-scale RCT where students with a 31–40\% predicted probability of completion were randomised to receive motivational contact, and subsequently, the study found significantly higher retention for the supported group. However, most institutional evaluations rely on observational data, where selection bias is a primary concern. \citet{de2024evaluating} recently addressed this by applying PSM to an early alert program, finding positive short-term impacts on student performance. The reliance on observational comparisons has motivated a shift toward more rigorous quasi-experimental designs to approximate randomised comparisons, particularly for studying AI-enabled outreach where random assignment is often not feasible \citep{de2024evaluating}.

\subsection{Propensity Score Matching in Educational Research}

To address the selection bias inherent in observational educational data, researchers increasingly rely on PSM to estimate the causal effects of interventions \citep{powell2020propensity}. In non-experimental settings where students cannot be randomly assigned to support services, PSM mimics the conditions of an RCT by creating comparable groups based on a ``propensity score'', which is the probability of receiving the treatment given a set of observed pre-treatment covariates. By matching treated and untreated units with similar propensity scores, researchers construct a control group that balances observed characteristics, thereby isolating the effect of the intervention from pre-existing differences in motivation, prior attainment, or demographics \citep{rosenbaum1983central}.
The utility of this approach is well-documented. A recent meta-analysis by \citet{surin2024quantifying} demonstrated that PSM is essential for unraveling the ``true effect sizes" of learning models, often revealing that raw observational differences overestimate intervention impacts. Consequently, PSM has become a standard tool for evaluating adaptive learning platforms and LMSs, allowing researchers to control for the self-selection bias often present when students voluntarily adopt digital tools \citep{lim2021changes, alalawi2025evaluating}.

\subsection{Applications and Methodological Evolution in Learning Analytics}

Propensity score methods have been applied across a wide range of educational settings, from evaluating teacher effectiveness \citep{alcott2017does} to special education services \citep{morgan2010propensity}. However, recent scholarship in LA has moved beyond basic matching techniques to employ more robust estimators, driven by the recognition that naive matching can be sensitive to algorithm choice \citep{wojciechowski2024intervention}.

\citet{mojarad2021replicating} compared PSM with Inverse Probability of Treatment Weighting (IPTW) in evaluating adaptive learning, suggesting that weighting methods can offer superior covariate balance and more stable estimates than simple 1:1 matching. Furthermore, to protect against model misspecification, advanced evaluations are increasingly adopting Doubly Robust Estimation (DRE), which combines propensity score weighting with outcome regression. For example, \citet{bowman2023does} used doubly robust analyses to confirm the effectiveness of supplemental instruction on retention, and \citet{wong2025assessing} applied DRE to quantify the impact of a university transition course, reporting a notable increase in pass rates. These studies illustrate a methodological maturation in the field, moving from simple descriptive comparisons to rigorous causal inference designs that account for the complex, hierarchical nature of educational data.
Table~\ref{tab:psm-studies} summarises representative examples of recent PSM-based evaluations in LA, spanning dashboards, advising agents, and generative AI assistants. These studies collectively highlight the necessity of transparent reporting regarding matching algorithms, covariate selection, and balance diagnostics.

\begin{table}[htbp]
\centering
\fontsize{8pt}{10pt}\selectfont
\begin{tabular}{p{1.0cm} p{1.5cm} p{2.3cm} p{2.1cm} p{2.2cm} p{1.7cm} p{3.6cm} }
\toprule
Context & Treatment / intervention & Sample size (analytic) & Covariates in propensity model & Matching method & Outcome(s) & Main result (post‐matching)  \\
\midrule
\textbf{University of Newcastle} \citep{alalawi2025evaluating} &
Student Progress and Performance Analytics (SPPA) learning analytics framework &
2019 control cohort $n=236$; 2021 SPPA cohort $n=248$; PSM‐matched sample $n=248$ per group (total $n=496$) &
Gender, program entry score (admission rank), age at census, citizenship / residency and related intake characteristics &
One-to-one nearest‐neighbour propensity score matching (logistic regression for PS; matching without replacement) &
Course pass / fail / withdrawal status; mean course mark &
SPPA use associated with a significantly higher pass rate and lower fail rate compared with matched controls; withdrawal rates and mean course marks showed no statistically significant differences; effects were strongest for students with lower prior entry scores. 
 \\
\midrule
\textbf{Western Canadian University} \citep{mousavi2021assessing} &
SARA (Student Advice / Activity and Reflection Assistant) learning analytics agent providing personalised advice / feedback &
First-year biology course; $N \approx 1{,}026$ students (analytic sample reduced after matching; exact matched $N$ depends on matching specification) &
Broad set of pre-treatment variables including demographic characteristics and prior academic performance indicators routinely available in the course dataset (e.g.\ prior grades); full list reported in the article &
Several matching strategies compared, including propensity‐score based nearest‐neighbour matching and Mahalanobis distance matching &
Course performance (final grade / pass result) and related academic outcomes &
Study’s primary contribution is methodological: estimated treatment effects for SARA were small and sensitive to the matching specification; no robust, large positive effect on course grades was found across specifications, and the authors emphasise the importance of careful covariate selection and matching diagnostics. 
 \\
\midrule
\textbf{Los Angeles Pacific University (LAPU)} \citep{hanshaw2025experimental} &
Spark AI course assistant integrated into LMS (GenAI assistant providing course-specific, Socratic help) &
2{,}090 student–course combinations (1{,}338 unique students) across 99 courses; PSM‐matched sample of 229 Spark users and 229 non-users (pairs within the same course and term) &
Age, gender, ethnicity (matching done within the same course and term to control for instructor and course effects) &
One-to-one propensity score matching to create matched pairs (within course/term), followed by non-parametric tests on matched pairs &
Course GPA (grade points on 0–4 scale) at the student–course level &
After PSM, Spark users had significantly higher GPAs than matched non-users (Wilcoxon signed-rank $p \approx 0.00026$; mean GPA difference $\approx 0.38$ points); effect size small–to–moderate. Authors explicitly caution that the quasi-experimental design supports strong association but not definitive causation. 
 \\
\midrule
\textbf{Online-vs.-face-to-face} \citep{rogers2022comparing} &
Research mentor training (RMT): comparison of synchronous online vs face-to-face mentor training for research mentors &
Overall mentor sample $N \approx 800$–$1{,}300$; focal analytic sample constructed via PSM to balance online and face-to-face mentors (online group about $n\approx 150$; matched face-to-face comparison group of similar size) &
Mentor and context characteristics such as discipline / field, mentor career stage and role, prior mentoring experience / prior RMT participation, gender, race/ethnicity, and other available baseline characteristics &
Propensity score matching to create balanced online and face-to-face mentor groups (one-to-one nearest neighbour on PS) &
Mentor‐reported gains in mentoring skills, self-efficacy and satisfaction; mentee ratings of mentoring quality; mentee research-related outcomes &
After matching, synchronous online RMT produced outcomes statistically comparable to traditional face-to-face RMT on mentor self-reported gains and mentee outcomes; no meaningful differences in effectiveness, supporting online delivery of RMT at scale.
 \\
\bottomrule
\end{tabular}
\caption{Recent examples of propensity score–based evaluations in higher education and learning analytics.}
\label{tab:psm-studies}
\end{table}

\paragraph{Challenges with Propensity Score Matching}

The most fundamental limitation of PSM is its reliance on the conditional ignorability assumption, that all confounders relevant to treatment assignment and outcomes are observed and measured. As \citet{weidlich2022causal} argue using Directed Acyclic Graphs (DAGs), observational LA studies are prone to specific biases, including confounding and collider bias, which cannot be resolved by data volume alone. The propensity score is a balancing score for \textit{measured} variables, not a device that removes hidden bias; thus, even excellent balance diagnostics cannot rule out distortion from unobserved factors such as student motivation or external life pressures \citep{Stuart2010Matching}. Consequently, rigorous reviews recommend explicit sensitivity analyses to quantify \textit{how strong} an unobserved confounder would have to be to invalidate the reported results \citep{Austin2011Intro}.
Furthermore, higher education data are hierarchical, with students embedded in course sections and programmes. Single-level matching can yield biased estimates when cluster-level attributes affect both selection and outcomes. Methodological work by \citet{leite2021multilevel} and \citet{guo2022effective} demonstrated that strategies respecting this clustering, such as within-cluster matching (\emph{blocking}, as implemented here within qualification/programme and study mode) or multilevel propensity score models, are necessary to reduce bias in educational settings. Additionally, selective programmes often create weak common support, requiring researchers to diagnose overlap and potentially trim the sample or use overlap weights to avoid extrapolating to dissimilar populations \citep{KingNielsen2019PSM}.

\subsection{Gaps in existing research}

Despite the growing body of work on early warning analytics and on propensity score methods in education, relatively few studies combine AI‐based risk stratification with rigorous causal evaluation of the subsequent proactive human interventions \citep{herodotou2020can}. 
Additionally, the PSM literature in higher education often focuses on traditional programmes or policy changes rather than on AI‐guided, student support‐driven support and while recent studies have begun to employ doubly robust estimation \citep{wong2025assessing}, theoretical perspectives that foreground the socio‐technical nature of these systems, such as AT, are seldom integrated with such quantitative impact evaluations.
This study sought to address these gaps by examining the effectiveness of an AI‐guided, proactive support intervention for at‐risk undergraduates in a large higher‐education institution, using PSM methods, doubly robust estimation, and sensitivity analyses to estimate the association between support and a range of academic outcomes, while conditioning on AI predictive scores and rich pre‐treatment covariates. We then interpret these findings through a theoretical lens, positioning this work as a bridge between rigorous AI–guided causal evaluation and an AT account of how student support is organised and enacted within institutional structures.

\section{Methodology}
\label{sec:methodology}

\subsection{Intervention cohort and control dataset}

The intervention cohort comprised 1,859 at-risk students who received at least one support case between 2022 and 2024. Case management logs show that the activity scaled over time, with 212 recorded interventions in 2022, 848 in 2023, and 1,486 in 2024, reflecting multiple contacts for some students. Most students were supported in a single year and through a single recorded case: the number of interventions per student had a median of 1 (mean = 1.31, SD = 2.09), with 137 (7.4\%) receiving more than one intervention. A smaller subset of 78 students received multi-year support, and only two students were followed across three consecutive years, which limits the extent to which cumulative outcomes can be attributed to repeated exposure rather than a single-year intervention. The intervention consisted of targeted, student support-led outreach delivered via email, telephone, and SMS, ranging from light-touch behavioural nudges to active pastoral and academic guidance as well as referrals. Meanwhile, the potential controls were drawn from students enrolled in the same programmes and years who had no recorded intervention during the study window.

\subsection{AI-based risk model for Probability of Success}

The intervention was driven by an institutional AI system that produces a Probability of Success (PoS) score for each enrolled student. This score is generated by a supervised learning model developed in prior work \citep{susnjak2024beyond} and trained on 250k historical student records. The model predicts successful qualification completion and is used operationally by the student support to prioritise outreach. 
Among the covariates, the AI-generated PoS plays a special role since it compresses a wide range of dynamic LMS interaction signals and static demographic as well as academic history into a single probability that a student will successfully complete their programme. Given its richness, it serves as the primary triaging signal for outreach by student support. 
The PoS has high accuracy. Its out-of-sample performance is strong (F1 = 0.871, precision = 0.829, recall = 0.942, AUC = 0.891, accuracy = 0.824), indicating that the model detects at-risk students with high sensitivity while maintaining good discrimination. In this study, the resulting PoS score serves two roles: (1) it is the central risk signal guiding human intervention decisions, and (2) it enters the propensity score specification to ensure that treated students are matched to controls with comparable AI-assessed risk.

\begin{figure}[htbp]
  \centering
  \includegraphics[width=\linewidth]{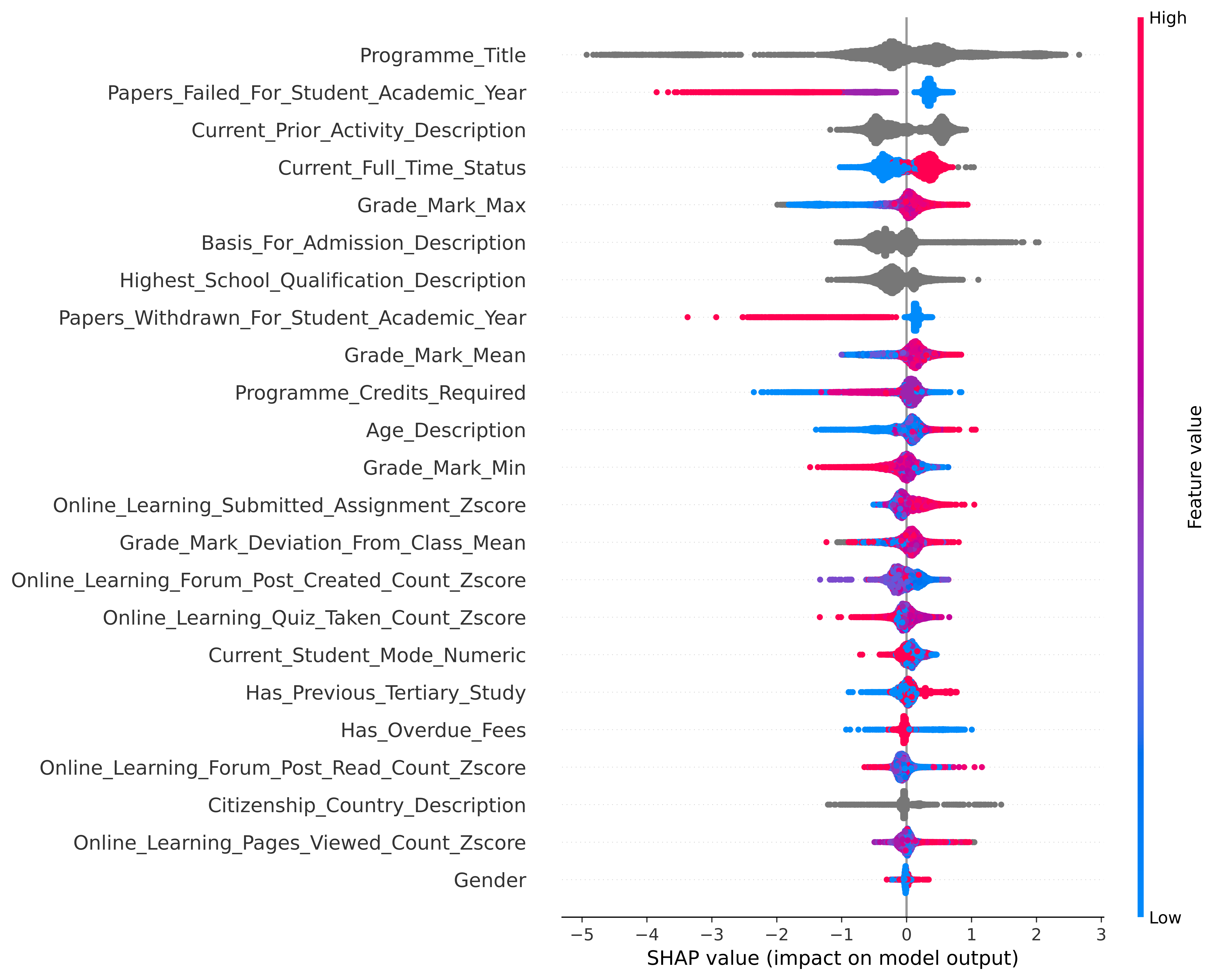}
  \caption{SHAP summary plot for the PoS model. Each point is a student record; position on the horizontal axis indicates the impact of the feature on predicted success, and colour indicates the feature value (blue = low, red = high). Features are ordered by overall importance.}
  \label{fig:shap}
\end{figure}

Model interpretability was examined using SHAP values. Figure~\ref{fig:shap} shows a SHAP summary (beeswarm) plot, where each point represents a student record, the horizontal axis gives the SHAP value (impact on predicted success), and colour encodes the feature value (blue = low, red = high). Features on the left are ordered by overall importance with respect to final predictions. Programme title, recent failures and withdrawals, prior activity status, enrolment mode and load, and grade history (mean, minimum, maximum, and deviation from class mean) emerge as the dominant drivers of risk, with higher failure counts and weaker grades pushing predictions towards qualification non-completion and stronger academic histories shifting them towards success. This provided assurance that the PoS model used for matching and targeting was anchored in pedagogically interpretable signals rather than opaque artefacts.

\subsection{Propensity score estimation and matching}
\label{sec:psm}

The propensity score and matching design involved first constructing a set of pre-treatment covariates that are frozen and represent a snapshot before any intervention decision, alongside an estimate for each student the probability of receiving support given these covariates, from which a matched comparison group of untreated students with similar profiles was formed. This design was strengthened by combining several elements: temporal alignment of covariates so that they are strictly pre-intervention, \textit{blocking} on programme/qualification and study mode so that students are only compared within comparable institutional contexts, a caliper on the logit of the propensity score so that matches are restricted to areas of good overlap, and a similarity metric that pays particular attention to the AI-derived PoS risk score while still balancing the broader covariate set. Here, blocking means forming strata defined by degree programme and study mode before any matching, while the caliper sets a maximum allowed difference in students’ log-odds of treatment (their logit propensity score) for them to be considered potential matches, and the logit transformation simply expresses the propensity score on the log-odds scale where distance is more informative for matching.

\paragraph{Pre-treatment covariates and temporal alignment}

The interventions were carried out by a central student support team, which initiated contact with
students identified as being at elevated risk of qualification non-completion.
To preserve a clear temporal ordering between predictors, treatment, and outcomes, all covariates entering the propensity score model and the matching design were defined using information that was available \emph{prior} to the intervention decision in a given academic year.
For academic performance, we constructed a lagged baseline snapshot at the end of the preceding calendar year (31 December of year $t-1$) and carried these values forward as pre-treatment covariates for interventions occurring in year $t$.
This snapshot included the variables in Table \label{tab:indicators}. 

\begin{table}[htbp]
\centering
\fontsize{8pt}{10pt}\selectfont
\caption{Baseline academic indicators and pre-intervention demographic, entry, and programme-context variables.}
\label{tab:indicators}
\begin{tabular}{p{1.2cm} p{5.9cm} p{7.3cm}}

\toprule
Domain & Concept & Example values \\
\midrule
\multirow{10}{1.2cm}{Baseline academic}  & Previous cumulative mean grade &
integer e.g., 58\%, 72\%, 85\% \\

 & Failed courses in previous academic year &
integer e.g., 0, 1, 3 \\

 & Withdrawn courses in previous academic year &
integer e.g., 0, 2 \\

 & Passed courses in previous academic year &
integer e.g., 2, 4, 6 \\

 & Cumulative courses attempted up to end of year $t-1$ &
integer e.g., 5, 12, 20 \\

 & Cumulative courses passed up to end of year $t-1$ &
integer e.g., 5, 10, 18 \\

 & Cumulative credits passed up to end of year $t-1$ &
integer e.g., 15, 120  \\

 & Programme-route mean grade in previous year &
integer e.g., 63\%, 70\% \\

 & Programme-level mean grade in previous year &
integer e.g., 65\%, 74\% \\

 & Prior qualification progression (percent completed) &
integer e.g., 40\%, 75\%, 100\% \\
\midrule

\multirow{6}{1.2cm}{Demographic / entry} & Age &
integer e.g., 19, 24, 37 \\

 & Gender &
Female, Male, Non-binary \\

 & Highest school qualification &
e.g., NCEA Level 3, Cambridge A-Levels \\

 & Basis for admission &
e.g., School leaver, Adult entry, International \\

 & First-in-family indicator &
 Yes / No \\

 & Years since last course &
integer e.g., 0, 1, 4, 10 \\

\midrule

\multirow{6}{1.2cm}{Study mode / workload} & Current study mode (distance / internal) &
Distance, Internal, Unknown \\

 & Current full-time status &
Full-time, Part-time, Unknown \\

 & Programme credits required &
integer e.g., 120, 360  \\

 & Enrolment length &
integer e.g., 1, 3, 6 \\

 & Hours per week on outside commitments &
integer e.g., 5, 20, 35 hours \\

 & Overdue fees count &
integer e.g., 0, 1, 3 \\
\midrule

\multirow{3}{1.2cm}{Programme/ progression} & Student journey stage &
early , middle, late \\

 & New to Massey status (new / returning) &
New or Returning \\

 & Programme title &
e.g., BSc Computer Science, BA Psychology, BBus Accounting \\
\bottomrule
\end{tabular}
\end{table}

These baseline academic indicators were combined with demographic, entry, and programme-context variables, all measured before the intervention year. Finally, we incorporated the AI-derived risk score, PoS score, as a pre-treatment covariate. While the PoS score is generated daily, for the analysis the score was aggregated on a rolling monthly basis for added stability. For each student--year, we aligned the PoS score to the month preceding any recorded intervention in that year and used this aligned value as a summary measure of pre-intervention risk. Only PoS values computed strictly before the intervention decision were used in the design to avoid any leakage of post-treatment information.

A critical methodological challenge in observational studies of student support is ``immortal time bias,'' particularly arising from the phenomenon of silent withdrawal or ``ghosting,'' where students disengage academically without formally notifying the institution. In such cases, bias can occur if the analysis compares treated students—who are by definition active at the time of intervention, with control students who may have already abandoned their studies weeks or months prior (perhaps longer), yet remain administratively enrolled. To mitigate this, our design leveraged the high-frequency temporal granularity of the AI risk model. As illustrated in the SHAP summary (Figure~\ref{fig:shap}), the PoS model relies heavily on dynamic behavioural features with the prefix \texttt{online\_} (e.g., 
\texttt{Online\_Learning\_Pages\_Viewed\_Count\_Zscore},\texttt{Online\_Learning\_Submitted\_Assignment\_Zscore}, \texttt{Online\_Learning\_Quiz\_Taken\_Count\_Zscore}), and these features capture real-time signals of student interaction with the LMS on a \textit{daily basis}. Consequently, a control student who had silently abandoned their studies prior to the intervention window would exhibit a precipitous decline in these signals, resulting in a distinctively low PoS score. By matching strongly on the PoS score (discussed subsequently) aligned to the month immediately preceding the intervention, we ensured that comparisons were made between students with equivalent levels of recent digital engagement and estimated \textit{survival probability}, effectively filtering out ``already-withdrawn/abandoned'' students from the potential control pool.

\paragraph{Propensity score model and common-support restriction}

Let $T_i \in \{0,1\}$ denote whether student $i$ received at least one support intervention in the focal year, and let $X_i$ collect the pre-treatment covariates listed above, including the aligned PoS score. We estimated each student’s propensity score
\[
e(X_i) = \Pr(T_i = 1 \mid X_i)
\]
using a logistic regression model with main effects for all covariates and pre-specified interactions between key programme-context and academic-history variables.
The propensity scores were then clipped to lie within $(\varepsilon, 1-\varepsilon)$, with $\varepsilon$ set to a small constant, to prevent extremely large inverse-probability weights in subsequent analyses.
Following standard practice, we restricted the analytic sample to the region of common support in the estimated propensity score.
Let $e_t$ and $e_c$ denote the estimated scores for treated and untreated students, respectively.
We computed
\[
\ell = \max\{\min(e_t), \min(e_c)\}, \qquad
h = \min\{\max(e_t), \max(e_c)\},
\]
and retained only students with $e(X_i) \in [\ell, h]$.
All matching and outcome analyses were conducted on this trimmed sample.

\paragraph{Blocking and caliper on the logit of the propensity score}

To respect the hierarchical structure of the data and to avoid matching across substantively dissimilar programmes, we defined a blocking factor by concatenating programme and study mode:
\[
\texttt{\_block}_i = \texttt{Programme Title}_i \,\Vert\, \texttt{Student Study Mode}_i,
\]
and performed matching only within blocks that contained at least one treated and one untreated student.
Within each block, we imposed a caliper on the logit of the propensity score, following recommendations such as those of \citet{Austin2011Intro}.
For each student in the common-support sample we defined
\[
\text{logit}(e(X_i)) = \log\left(\frac{e(X_i)}{1 - e(X_i)}\right),
\]
computed the standard deviation of $\text{logit}(e(X_i))$ in the trimmed data, and set the caliper width to $0.2$ times this standard deviation.
A control student $j$ was considered a candidate match for treated student $i$ only if
\[
\bigl|\text{logit}(e(X_i)) - \text{logit}(e(X_j))\bigr| \leq \text{caliper}.
\]
Treated students with no eligible controls under this restriction were excluded from the matched sample.

\paragraph{Distance metric and use of the AI risk score in matching}

Among the set of candidate controls within the same block and caliper window, we measured similarity using Gower distance, which averages variable-wise dissimilarities and is appropriate for mixed numeric and categorical data \citep{gower1971general}.
Given that PoS is a strong prognostic summary of academic risk and was central to the institution’s decision process, we did not treat it as just another covariate in the Gower distance.
Instead, we constructed a composite distance that optimised matching on PoS while still enforcing close similarity on the broader covariate set.
First, PoS was rescaled to the unit interval within the trimmed sample.
Then, for a treated--control pair $(i,j)$ we defined
\[
d_{ij}
= (1 - w)\, d_{\text{Gower}}(X_i, X_j)
  + w\, \bigl|\widetilde{\text{PoS}}_i - \widetilde{\text{PoS}}_j\bigr|,
\]
where $d_{\text{Gower}}(\cdot,\cdot)$ is the Gower distance on the covariates above,
$\widetilde{\text{PoS}}$ denotes the rescaled risk score, and $w$ is a tuning parameter controlling the relative importance of PoS.
The propensity score model thus included PoS together with demographic, enrolment,
and prior performance covariates, and the composite distance in our methodological design was set to give it dominant weight ($w = 0.75$) to ensure effective filtering out of students who had already disengaged to mitigate the immortal time bias, while still enforcing good alignment on the remaining
pre-treatment characteristics.

\paragraph{Matching ratio and construction of matched pairs}

Within each block, and subject to the logit caliper, we performed nearest-neighbour matching on the composite distance $d_{ij}$.
For each treated student we selected up to $m=2$ control students with the smallest distance among eligible candidates, allowing controls to be reused across different treated students (matching with replacement).
Matching with replacement reduces the risk of poor matches in blocks with relatively few suitable controls and supports better covariate balance at the cost of increased variance \citep{Austin2011Intro}.
The result is a matched sample organised as treated--control pairs (or $1{:}2$ sets) with a MultiIndex structure over columns, where each row contains the pre-treatment covariates and outcomes for one treated student and its corresponding matched control student(s).
All subsequent outcome modelling, weighting, and doubly robust estimation were carried out on this matched sample.

\subsection{Outcome modelling and doubly robust estimation}
\label{sec:dr}

Having constructed a matched sample in which treated and untreated students are comparable on observed pre-treatment characteristics, the second stage of the analysis estimated the effect of support on subsequent academic outcomes. Rather than relying only on post-match mean differences, we used a doubly robust framework that combines outcome regression with inverse-probability weighting based on the estimated propensity scores. This approach produces an ATE estimate that remains consistent if either the propensity score model or the outcome regression models are correctly specified, and it also stabilises the influence of units with very low or high treatment probabilities.
The primary causal estimand was the ATE of receiving a targeted support intervention on a set of academic outcomes measured at the end of the intervention year. All outcome analyses were conducted on the matched sample described in Section~\ref{sec:psm}, using only post-treatment outcomes and pre-treatment covariates.

\paragraph{Family selection and arm-specific outcome models}

For each outcome variable $Y$ in the analysis set (annual and cumulative counts of courses and
credits, percentages of courses passed, grade averages, and qualification progression measures),
we estimated separate outcome regression models for the treated and control arms, adjusting
for the same pre-treatment covariates used in the propensity score design.

Let $Y^{(1)}$ and $Y^{(0)}$ denote the post-treatment outcomes for treated and matched control
students respectively, and let $X$ denote the matrix of pre-treatment covariates after
binary encoding, aligned row-wise with the matched pairs.
For each outcome, we first selected an appropriate regression family in a data-driven way.
If the treated-arm outcome values were non-negative and approximately integer-valued, we treated the outcome as a count and used a Poisson generalised linear model; otherwise we used a Gaussian linear model.
Formally, for family $\mathcal{F} \in \{\text{Poisson}, \text{Gaussian}\}$ we fit
\[
Y^{(1)} \mid X \sim \mathcal{F}(\eta^{(1)} = \alpha^{(1)} + X\beta^{(1)}), \qquad
Y^{(0)} \mid X \sim \mathcal{F}(\eta^{(0)} = \alpha^{(0)} + X\beta^{(0)}),
\]
where $\eta$ is the linear predictor.
All models included an intercept and were estimated with heteroskedasticity-robust
(HC0) sandwich standard errors.
For Poisson outcomes, we used iteratively reweighted least squares; if the Poisson fit
encountered numerical instability for a given outcome, the code automatically fell back
to a robust OLS specification with the same predictors.
From these fitted models we obtained arm-specific predicted outcomes
$\hat\mu^{(1)}_i = \hat{\mathbb{E}}[Y^{(1)}_i \mid X_i]$ and
$\hat\mu^{(0)}_i = \hat{\mathbb{E}}[Y^{(0)}_i \mid X_i]$ at the covariate values $X_i$
of each matched pair.

\paragraph{Stabilised propensity weights and doubly robust estimator}

To construct a doubly robust ATE, we combined the arm-specific outcome models with the
previously estimated propensity scores.
For each student $i$ in the matched sample, we extracted the corresponding propensity score
$\hat e_i$, either from arm-specific columns in the matched data or, if unavailable, from a
single score vector aligned to rows.
Propensity scores were trimmed to lie within $[0.10, 0.90]$ in the main analysis to avoid
extreme inverse probabilities, and stabilised inverse-probability weights were computed as
\[
w^{(1)}_i = \frac{\pi}{\hat e_i}, \qquad
w^{(0)}_i = \frac{1-\pi}{1-\hat e_i},
\]
where $\pi$ is the marginal treatment rate in the matched sample.
The stabilisation by $\pi$ preserves the expectation of the doubly robust estimator
while reducing the variance contribution from very small or very large propensity
scores.

Using these components, we formed augmented inverse-probability influence functions for
each observation:
\[
\phi^{(1)}_i = \hat\mu^{(1)}_i + w^{(1)}_i\bigl(Y^{(1)}_i - \hat\mu^{(1)}_i\bigr), \qquad
\phi^{(0)}_i = \hat\mu^{(0)}_i + w^{(0)}_i\bigl(Y^{(0)}_i - \hat\mu^{(0)}_i\bigr),
\]
and defined the doubly robust ATE estimator as the sample mean of the contrast
\[
\widehat{\text{ATE}}_{\text{DR}}
= \frac{1}{n}\sum_{i=1}^n \bigl(\phi^{(1)}_i - \phi^{(0)}_i\bigr).
\]
This estimator is consistent for the ATE under standard assumptions if either the propensity
score model or both arm-specific outcome models are correctly specified.
For added robustness to extreme residuals, we optionally winsorised the per-observation
contrasts $\phi^{(1)}_i - \phi^{(0)}_i$ at the 1st and 99th percentiles before averaging;
the main results reported here used this winsorisation for sensitivity checks but were
qualitatively unchanged without it.

\paragraph{Bootstrap uncertainty quantification and cross-checks}

Uncertainty for $\widehat{\text{ATE}}_{\text{DR}}$ was quantified via nonparametric
bootstrap at the level of matched rows.
For each outcome we generated $B=999$ bootstrap resamples by drawing, with replacement,
$n$ indices from $\{1,\dots,n\}$ and recomputing the mean of the corresponding
influence-function contrasts.
The empirical 2.5th and 97.5th percentiles of the bootstrap distribution provided
a percentile-based 95\% confidence interval, and a two-sided bootstrap $p$-value was
computed as twice the smaller of the proportions of bootstrap replicates above or below zero.

To facilitate diagnostic comparisons and to guard against implementation errors, we also
computed two additional effect estimates for each outcome:

\begin{itemize}
    \item the simple post-treatment mean difference in the matched sample,
    \[
    \widehat{\Delta}_{\text{post}} = \frac{1}{n}\sum_{i=1}^n \bigl(Y^{(1)}_i - Y^{(0)}_i\bigr),
    \]
    which ignores covariates but respects the matching structure; and
    \item an ANCOVA-style regression estimate obtained by pooling treated and control
    observations, regressing the post-treatment outcome on a treatment indicator and the
    pre-treatment covariates, and extracting the coefficient on the treatment indicator.
\end{itemize}

For each outcome we also recorded summary diagnostics on the effective propensity weights,
including the minimum and maximum trimmed propensity scores in each arm and the 99th
percentiles of the stabilised weights for treated and control units.
These diagnostics confirmed that the trimming removed extreme propensity scores and that
the weight distributions were well-behaved, thereby supporting the numerical stability of
the doubly robust estimator.
Finally, the above procedure was applied uniformly to the full set of pre-specified outcomes
(annual and cumulative performance measures and qualification progression), so that each
effect estimate was based on the same matched sample, the same set of pre-treatment covariates, and an outcome-specific but principled choice of regression family.

\paragraph{Covariate balance and sensitivity to hidden bias}

After constructing the matched sample, we formally assessed covariate balance between
treated and control students and examined the robustness of the main effects to
unobserved confounding.
First, we computed standardised mean differences (SMDs) for all pre-treatment
covariates used in the propensity score design, extracting the corresponding treated and control series and classifying them as numeric or categorical.
Numeric covariates were compared using the usual pooled-variance SMD based on means
and variances in the two arms.
Categorical covariates were one-hot encoded into level indicators, and SMDs were
computed for each level’s treated–control proportion difference, using a pooled
binomial variance.
We computed per-level SMDs and summarised the maximum absolute SMD for each covariate to verify that post-matching balance met conventional thresholds. Matching greatly reduced pre-existing differences where all covariates had $|SMD| < 0.25$ and most, including PoS and key prior performance indicators, had $|SMD| < 0.10$, with the remaining moderate imbalance confined to structural variables such as age and enrolment length.

Second, for selected primary outcomes we carried out a Rosenbaum-type sensitivity
analysis for hidden bias on the matched pairs.
For each outcome, we formed paired differences between treated and control students
and computed the Wilcoxon signed-rank statistic, restricting to non-zero differences.
Under the Rosenbaum model with sensitivity parameter $\Gamma \ge 1$, we then derived
upper and lower bounds on the one-sided $p$-value for a sequence of $\Gamma$ values
(e.g., $\Gamma \in \{1.0, 1.25, 1.5, 2.0, 2.5\}$), using the large-sample normal
approximation to the signed-rank statistic with bias-adjusted mean and variance.
This yielded, for each outcome, the range of $p$-values that would be compatible with
a given degree of unobserved bias in the odds of treatment, and allows us to report
the largest $\Gamma$ at which the observed effect remains statistically significant.

In addition, we implemented a more robust balance-checking routine to guard against
edge cases in the matched data structure.
Because the matched dataset could contain duplicate column names arising from
earlier merges, we first collapsed any duplicate covariate columns within each arm by
taking the first non-missing value row-wise, ensuring that each covariate was
represented by a single well-defined series.
We then classified covariates as numeric or categorical using a data-driven rule based
on the proportion of values that could be reliably coerced to numeric form and the
resulting cardinality.
Numeric covariates with sufficient numeric content and at least a minimum number of
distinct values were treated as continuous and compared via pooled-variance SMDs,
whereas remaining covariates were treated as categorical, converted to category type,
and expanded into one-hot indicators for level-specific SMDs.
From these, we derived both a long-format table of per-level SMDs and a summary of the
maximum absolute SMD per covariate, which we used as our main diagnostic of post-match
balance.

We also refined the sensitivity analysis to focus on outcomes for which the doubly robust
ATE indicated statistically significant effects.
Starting from the DR results table (containing each outcome’s $\widehat{\text{ATE}}_{\text{DR}}$ and bootstrap $p$-value), we filtered to outcomes with $p_{\text{boot}} < 0.05$ and, for each, set the direction of the Rosenbaum Wilcoxon test to match the sign of the estimated effect (``greater'' for beneficial increases such as higher grades or pass percentages, ``less'' for beneficial decreases such as failure counts).
For each such outcome we then computed Rosenbaum Wilcoxon bounds over a grid of
$\Gamma$ values, reporting the corresponding upper and lower $p$-value bounds and the
number of informative matched pairs.
This procedure ensures that the hidden-bias sensitivity analysis is aligned with the
primary causal claims and is applied consistently only to those outcomes where a
meaningful effect is estimated.

\subsection{Tooling}
All statistical analyses were conducted using Python and various libraries. The \texttt{scikit-learn} library was employed for propensity score estimation, and the \texttt{gower} library was used to compute distance metrics for matching. Doubly robust estimation was performed using custom implementations that relied on \texttt{statsmodels} for the underlying outcome regression models.

\section{Results}
\label{sec:results}

\subsection{Post‐match balance on AI risk score}

\begin{figure}[t]
  \centering
  \includegraphics[width=\columnwidth]{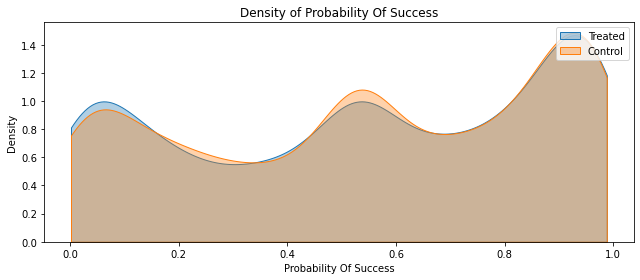}
  \caption{Density of AI probability of success (PoS) for treated and matched control students after propensity score matching.}
  \label{fig:pos_density}
\end{figure}

\begin{figure}[htbp]
  \centering
  \begin{subfigure}[b]{0.48\textwidth}
    \centering
    \includegraphics[width=\textwidth]{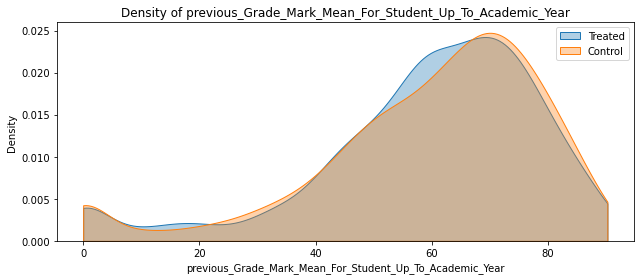}
    \caption{Cumulative mean grade up to the academic year prior to intervention.}
    \label{fig:mean_grade_prev}
  \end{subfigure}\hfill
  \begin{subfigure}[b]{0.48\textwidth}
    \centering
    \includegraphics[width=\textwidth]{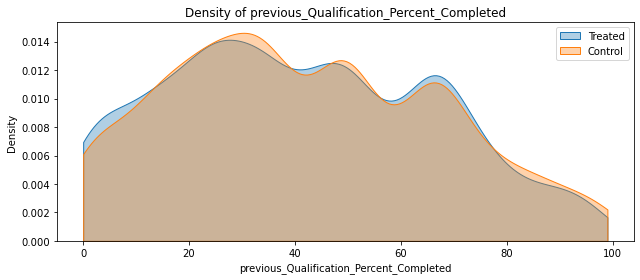}
    \caption{Percent of qualification completed in the year prior to intervention.}
    \label{fig:qualification_percent_up_to_year}
  \end{subfigure}
  \caption{Selection of diagnostic density plots on key covariates between treated and control students.}
    \label{fig:diagnosti_density1}
\end{figure}

Figure~\ref{fig:pos_density} shows the distribution of the AI risk score, PoS score, for treated students and their matched controls in the final analytic sample.
The two density curves almost completely overlap across the full range from 0 to 1, indicating that the matching procedure successfully aligned treated and control students on the key pre‐treatment signal based on a highly calibrated machine learning model that guided human support decisions.
Given that the student support team explicitly targeted students with elevated risk according to PoS, this alignment is important because it suggests that the estimated treatment effects compare students who were similarly assessed by the institutional AI system as being at-risk of non‐completion.
The distributions of two further covariates are shown in Figure \ref{fig:diagnosti_density1}, highlighting robust matching for grade score and qualification percentages completed up to the year-end immediately before the year of the intervention.
The post-match balance between the remaining covariates of the treated and the control pairs was also robust and is shown in the Appendix \ref{appdx:psm-diagnostics}.

\subsection{Effects on academic performance in the intervention year}

Table~\ref{tab:dr_results} summarises the average treatment effects estimated using the doubly robust procedure.
The first block focuses on performance as recorded by the end of the intervention year.
The intervention is associated with a reduction in the number of failed courses. 
On average, treated students failed around three quarters ($-0.78$) of a course fewer than comparable controls by the end of the year in which they received support. In real terms, this equates to avoiding 78 course failures for every 100 students supported.
Other year-end outcomes move in a direction that is consistent with this positive pattern, but their confidence intervals include zero; thus, introducing a degree of uncertainty.
The intervention is estimated to reduce total course withdrawals in the treatment year by about a fifth of courses on average for treated students, and to increase the number of courses passed by about 0.6 on average. This translates to an annual percentage of courses passed to about 12.8 percentage points higher than that of control students, alongside 10.5 extra credits earned. Given the proximity of the confidence intervals to zero values, while not conclusive, they are suggestive of positive effects.

\begin{table}[htbp]
\centering
\caption{Doubly robust estimates of the average treatment effect (ATE\(_{\text{DR}}\)) for annual and cumulative academic outcomes. Negative values for failure- and withdrawal-type outcomes indicate better results for supported students.}
\label{tab:dr_results}
\small
\begin{tabular}{llccc}
\toprule
Domain & Outcome & $\widehat{\text{ATE}}_{\text{DR}}$ & 95\% CI & $p_{\text{boot}}$ \\
\midrule
\multicolumn{5}{l}{\textit{Intervention year outcomes}} \\
& Courses failed in intervention year
  & $-0.78$ & $[-1.34,\,-0.17]$ & $0.018$ \\
& Courses withdrawn in intervention year
  & $-0.19$ & $[-0.47,\,0.11]$  & $0.188$ \\
& Courses passed in intervention year
  & $0.61$  & $[-0.45,\,1.60]$  & $0.258$ \\
& Percent of courses passed in intervention year
  & $12.82$ & $[-1.60,\,26.61]$ & $0.072$ \\
& Credits passed in intervention year
  & $10.49$ & $[-6.05,\,26.24]$ & $0.214$ \\
\midrule
\multicolumn{5}{l}{\textit{Cumulative outcomes up to end of intervention year}} \\
& Cumulative mean grade
  & $8.82$  & $[2.74,\,13.97]$  & $0.008$ \\
& Cumulative deviation from class mean
  & $7.43$  & $[2.43,\,11.88]$  & $0.004$ \\

& Cumulative courses passed & $1.85$ & $[-0.72,\,4.45]$ & $0.162$ \\ 
& Cumulative courses failed & $-3.60$ & $[-5.31,\,-1.92]$ & $<0.001$ \\ 
& Cumulative courses withdrawn & $-0.56$ & $[-1.44,\,0.34]$ & $0.232$ \\ 
& Cumulative credits passed & $30.78$ & $[-8.55,\,69.69]$ & $0.138$ \\ 
& Cumulative percent of courses passed & $13.99$ & $[4.68,\,22.88]$ & $0.004$ \\
  
& Qualification percent completed
  & $7.94$  & $[-1.38,\,17.62]$ & $0.116$ \\
\bottomrule
\end{tabular}

\vspace{0.3em}
\begin{flushleft}
\footnotesize
Note. ATE\(_{\text{DR}}\) is the doubly robust average treatment effect (treated minus control) from the matched sample with stabilised propensity weights. Confidence intervals are percentile bootstrap intervals. Negative values for failure and withdrawal outcomes indicate fewer adverse events for supported students; positive values for pass rates, credits, grades, and completion indicate better outcomes for supported students.
\end{flushleft}
\end{table}

\subsection{Cumulative academic outcomes and qualification progression}

The cumulative outcomes, which aggregate performance up to the end of the intervention year, show a clear pattern of benefit, particularly with respect to grades.
The intervention is associated with higher cumulative grades.
The estimated effect is that treated students will attain mean grades that are, on average, 8.8 points higher than control students, with a confidence interval from 2.7 to 14.
When grades are expressed relative to class peers and thus normalised to relevant courses in the form of mean grade deviations from class averages, the effect remains positive and large, with treated students performing on average 7.4 points better than control students, having a confidence interval from 2.4 to 11.9.
These findings indicate that supported students not only raise their absolute performance but also improve their standing within their enrolled courses compared to their equivalent peers.
Beyond grades, the cumulative volume measures indicate that by the end of the intervention year, supported students had a more favourable transcript history than comparable controls. The total number of failed courses accumulated up to that point was lower by an estimated $3.6$ courses on average, with a 95\% confidence interval from $-5.3$ to $-1.9$, which implies that even after conditioning on their weaker pre–intervention histories, supported students experienced materially fewer failures across their enrolment. Consistent with this, the cumulative percentage of courses passed was higher by about $14$ percentage points, indicating that a substantially larger share of their enrolment was successfully completed. The cumulative numbers of courses passed and credits earned were also higher for supported students (1.9 courses and 30.8 credits, respectively), while cumulative withdrawals were lower by about half a paper, although the confidence intervals for these three outcomes include zero. When considered together with the annual outcomes, these cumulative indicators suggest that the intervention not only improved performance during the treatment year but also contributed to a more successful overall progression trajectory by the end of that year, even when accounting for students’ prior academic records.
Similarly, the overall qualification progression measured by `qualification percent completed' shows an estimated gain of about 7.9 percentage points over control students (albeit with a confidence interval that includes zero).
This outcome agrees with courses passed and credits earned in the intervention year, and together strongly hint at improved overall progression.
Thus, the patterns from cumulative outcome variables support the substantive claim that the decision-support from the institutional AI and the student support team correctly targeted students who were on a worse trajectory and that, conditional on this predicted higher risk, the intervention improved their cumulative outcomes.

\paragraph{Weight diagnostics and robustness}

Across outcomes, the stabilised propensity weights showed moderate dispersion but no signs of extreme instability.
The 99th percentile of the treated weights was approximately 10, while the corresponding value for the control weights was about \textasciitilde1.3.
This indicates that a small fraction of treated students received relatively high weights, which is expected in designs that rely on matching with replacement and trimming of propensity scores, but that no single observation dominated the estimates.
The agreement between the doubly robust estimates, the ANCOVA cross checks, and the expected directions of effect further supports the internal consistency of the analysis.

\paragraph{Sensitivity to unobserved confounding}

To assess the robustness of the main effects to potential unobserved confounding, we computed Rosenbaum Wilcoxon signed–rank sensitivity bounds on the matched pairs for six outcomes with statistically significant doubly robust ATEs. For each outcome we examined how the upper bound on the one–sided $p$–value changed as the sensitivity parameter $\Gamma$ increased, which represents the degree to which an unmeasured factor could change the odds of receiving the intervention among otherwise matched students.
Across these outcomes, the observed effects remained statistically significant for moderate values of $\Gamma$ (up to the largest value considered in our grid for some outcomes), indicating that an unobserved covariate would need to change the odds of treatment by a substantial margin within matched pairs in order to explain away the estimated benefits. 
For example, for courses failed in the intervention year, the effect remained significant up to $\Gamma \approx 2.5$, meaning that even if an unobserved  factor made some students about 150\% more likely to receive support than their matched counterparts, the conclusion would not change. Full Rosenbaum bounds across the main effects are reported in Appendix~\ref{app:rosenbaum}.

\section{Discussion}

\subsection{Interpreting AI--guided support through Activity Theory}

From an Activity Theory (AT) perspective, the intervention evaluated here operates as a complex socio–technical activity system. Rather than viewing the AI model merely as a predictive instrument, AT frames the PoS score as a critical \textit{mediating artefact}. This artefact transforms the abstract, invisible concept of ``statistical risk'' into a concrete, actionable signal that allows the \textit{subjects} (student support team) to direct their labour toward a shared \textit{object}, which is the stabilisation of precarious student trajectories \citep{Engestrom2015Learning}. The robust causal estimates provided by the doubly robust analysis offer more than just a measure of impact, additionally, they provide a window into the internal contradictions and resolutions within this system.

The quantitative findings reveal a distinct pattern. The intervention was highly effective at reducing failed courses and improving cumulative grades, yet its impact on the speed of qualification completion was positive but statistically uncertain. Through an AT lens, this dichotomy can be understood as a tension between the immediate \textit{pedagogical object} and the longer-term \textit{institutional object}. The reduction in failed courses indicates that the primary contradiction (the clash between a student’s risk profile and the academic demands of the intervention year) was successfully resolved by the student support–tool configuration. The AI model acted as a spotlight, and the student support provided the necessary scaffolding. In this sense, the activity system functioned effectively as a ``brake'' on failure, stopping the accumulation of negative outcomes that typically lead to withdrawal. The fact that treated students failed substantially fewer courses than their matched peers supports the notion that when the tool mediates the subject's action, the immediate academic crisis is mitigated.

However, the uncertainty regarding qualification completion highlights a secondary contradiction located between the \textit{outcome} of the intervention and the rigid \textit{rules} of the wider institution. While the student support–tool dyad successfully improved academic performance (higher grades, fewer failures), this recovery does not automatically translate into accelerated degree completion. The \textit{rules} governing higher education such as prerequisite chains, semester schedules, and maximum course loads, often mean that a student who is ``saved'' from failure must still traverse a fixed temporal path to graduate. The intervention improved the \textit{quality} of the student's journey (performance) but could not easily overcome the structural constraints determining the \textit{velocity} of that journey. This suggests that while the division of labour between AI and student support teams is optimised for retention, the broader activity system is not yet configured to capitalise on this stability to accelerate progression.

Furthermore, this study represents a second–order activity in which institutional researchers use PSM and sensitivity analyses as tools to interrogate the primary support activity \citep{Engestrom1999pat}. The analysis exposes a latent contradiction between the precision of the AI risk score and the bluntness of the intervention modality. The PoS model (the tool) is highly granular, updating daily, yet the intervention (the action) is often a single, binary event. The finding that cumulative gains in grades were significant suggests that even this limited ``dose'' of support can reorient a student's trajectory. However, the expansive learning cycle described by Engestr"om would suggest that to move the needle on qualification completion, the \textit{community} must renegotiate the \textit{rules} of engagement, perhaps moving from single-point interventions to continuous, adaptive support that mirrors the dynamic nature of the risk score.

Integrating AT into the interpretation of these causal findings; therefore, moves beyond descriptive mapping. It explains \textit{why} the intervention works the way it does, from which it can be deduced that it is a highly effective stabilisation mechanism (resolving immediate risk) operating within a rigid structural framework (constraining rapid completion). This framing links the individual–level treatment effects directly to institutional design choices, suggesting that sustaining the observed gains requires not just better algorithms, but a reconfiguration of how institutional rules and resources support the students who have been successfully stabilised by the intervention.

\subsection{Key Findings and implications}

Therefore, in circling back to the study's original questions, the answer to \textbf{RQ1}, the propensity score and doubly robust analyses show that AI--guided student support is associated with a clear reduction in course failures and improved cumulative grades and pass rates for at--risk students, while effects on qualification completion are positive but less precise. With respect to \textbf{RQ2}, the Activity Theory lens frames this intervention as a socio--technical system in which an AI risk model, student support staff, and institutional rules jointly mediate how risk is identified and addressed. The findings imply that the current configuration functions effectively as a stabilisation mechanism, preventing the escalation of academic risk, but that the translation of these gains into faster completion is constrained by programme structures and progression rules. Practically, this points to the need to pair predictive analytics with reforms to curriculum pathways and ongoing, multi-contact support models that align the granularity of intervention with the dynamic nature of the risk signal.

\subsection{Limitations and future work}

This study has several limitations that qualify the findings. First, the analysis relies on observational data and propensity score methods, so residual unobserved confounding cannot be ruled out even with doubly robust estimation and Rosenbaum sensitivity checks. Second, key predictors had to be constructed as lagged, end–of–year variables to avoid data leakage, which reduced temporal granularity and prevented us from modelling within–year dynamics or the timing of support relative to critical academic events. Third, because the propensity score model required prior achievement histories, some first–year students and others without sufficient enrolment history could not be matched and were excluded; as a result, the estimates primarily describe effects for continuing at–risk students rather than for the full population of new full-time entrants. Furthermore, while the inclusion of the time-aligned PoS score serves as a strong proxy for active enrolment to mitigate immortal time bias, it remains a probabilistic measure based on digital footprints; thus, some residual bias may persist for control students who maintained minimal LMS activity despite having psychologically disengaged from their studies. Finally, although the dataset included students who received multiple interventions of varying intensity, the analysis operationalised support (emails, calls, texts) as a single binary exposure. By aggregating heterogeneous contact modes and frequencies into one treatment indicator, the study effectively treats the support process as a ``black box.'' Consequently, we cannot disentangle whether the observed benefits were driven by a single timely nudge or by repeated engagement, nor can we identify which specific forms of support were most effective. Data acquisition regarding the nature of the interventions has now been improved and this is a topic for subsequent research. 

These constraints suggest several alternative directions for future research. Longer panels and finer–grained data would permit longitudinal causal designs (for example, marginal structural models or repeated–treatment PSM) to examine how multi–year patterns of AI–guided support shape trajectories and completion. Richer encoding of intervention modality, timing, and intensity would enable the study of dose–response relationships and heterogeneous treatment effects across programmes, demographics, and PoS risk bands. Combining the present quantitative framework with qualitative or log–level analyses within an Activity Theory lens could also clarify how student support teams’ practices, programme structures, and institutional rules mediate the impact of AI–driven alerts, and could support the design and simulation of alternative targeting and workload policies before widescale implementation.

\section{Conclusion}

This study embedded AI-generated risk scores from a predictive model, directly into propensity score matching for causal evaluation of proactive interventions, interpreted via Activity Theory to reveal socio-technical bottlenecks.
This findings showed that AI–guided academic support, when targeted using a well-calibrated risk model and evaluated with propensity score matching and doubly robust estimation, can meaningfully improve outcomes for at–risk tertiary students. Compared with matched peers, supported students accumulated fewer failed courses, achieved higher cumulative grades, and passed a larger share of their enrolled courses by the end of the intervention year, while effects on qualification completion were positive. These findings suggest that combining institutional AI with human student support can shift the trajectories of students who are already on a weaker path, rather than merely reflecting pre–existing differences. When viewed through an Activity Theory lens, the intervention can be framed as a socio–technical system in which AI tools, student support teams, and institutional rules jointly shape what support is delivered and to whom. The results indicate that the student support–tool configuration is functioning in a beneficial way, but also that programme structures and broader institutional constraints limit how quickly improved performance translates into completed qualifications. Future work should exploit finer–grained, multi–year data and richer intervention logs to trace how different modes and intensities of AI–guided support operate over time, and to inform the redesign of targeting thresholds, workloads, and governance so that the gains observed here can be strengthened and extended to a wider range of students.

\section*{Ethics Approval}
Ethical approval for this study was granted by the Massey University Human Ethics Committee (Application ID: HE014 – OM3 23/42). All data were processed and analyses conducted in accordance with the institutional guidelines and the Committee’s conditions for the use of student records in research.

\bibliographystyle{apalike}

\appendix


\section{PSM Diagnostics}
\label{appdx:psm-diagnostics}
--------------------------------------------------------

\begin{figure}[htbp]
  \centering
  \begin{subfigure}[b]{0.48\textwidth}
    \centering
    \includegraphics[width=\textwidth]{mean_grade.png}
    \caption{Previous mean grade up to academic year.}
    \label{fig:mean_grade_prev}
  \end{subfigure}\hfill
  \begin{subfigure}[b]{0.48\textwidth}
    \centering
    \includegraphics[width=\textwidth]{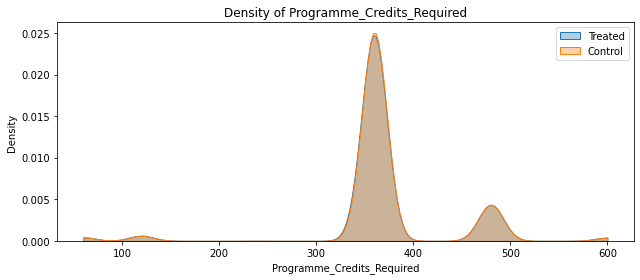}
    \caption{Programme credits required.}
    \label{fig:credits_required}
  \end{subfigure}
  \caption{Density plots for selected baseline covariates (row 1).}
  \label{fig:continuous_row1}
\end{figure}

\begin{figure}[htbp]
  \centering
  \begin{subfigure}[b]{0.48\textwidth}
    \centering
    \includegraphics[width=\textwidth]{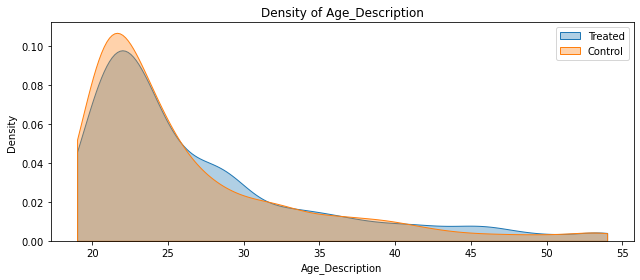}
    \caption{Student age.}
    \label{fig:age}
  \end{subfigure}\hfill
  \begin{subfigure}[b]{0.48\textwidth}
    \centering
    \includegraphics[width=\textwidth]{qualification_percent_up_to_year.png}
    \caption{Previous qualification percent completed.}
    \label{fig:qual_percent}
  \end{subfigure}
  \caption{Density plots for selected baseline covariates (row 2).}
\end{figure}

\begin{figure}[htbp]
  \centering
  \begin{subfigure}[b]{0.48\textwidth}
    \centering
    \includegraphics[width=\textwidth]{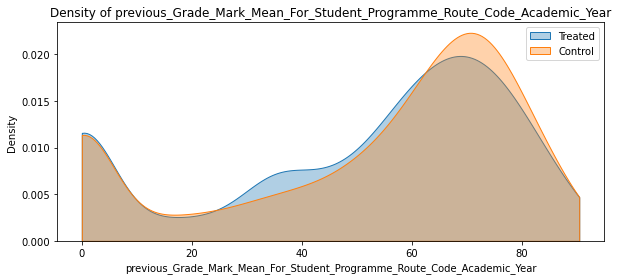}
    \caption{Previous programme-route mean grade.}
    \label{fig:mean_grade_route}
  \end{subfigure}\hfill
  \begin{subfigure}[b]{0.48\textwidth}
    \centering
    \includegraphics[width=\textwidth]{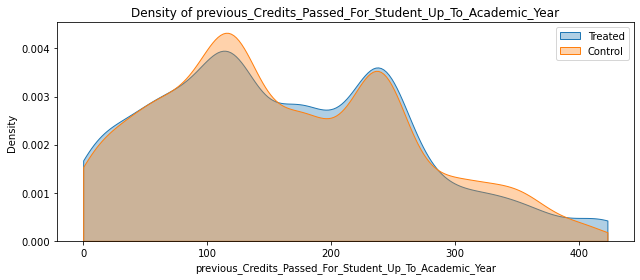}
    \caption{Previous total credits passed.}
    \label{fig:total_credits}
  \end{subfigure}
  \caption{Density plots for selected baseline covariates (row 3).}
\end{figure}

\begin{figure}[htbp]
  \centering
  \begin{subfigure}[b]{0.48\textwidth}
    \centering
    \includegraphics[width=\textwidth]{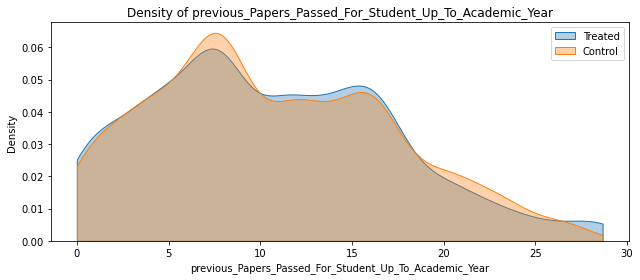}
    \caption{Previous total courses passed.}
    \label{fig:total_papers}
  \end{subfigure}\hfill
  \begin{subfigure}[b]{0.48\textwidth}
    \centering
    \includegraphics[width=\textwidth]{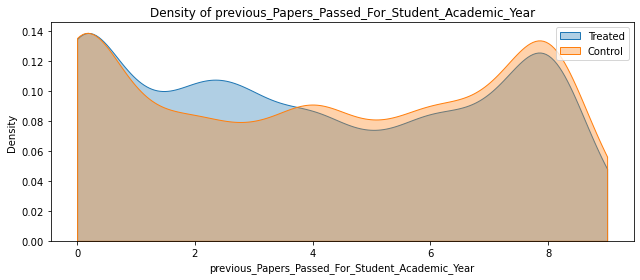}
    \caption{Previous courses passed (last academic year).}
    \label{fig:papers_passed}
  \end{subfigure}
  \caption{Density plots for selected baseline covariates (row 4).}
\end{figure}

\begin{figure}[htbp]
  \centering
  \begin{subfigure}[b]{0.48\textwidth}
    \centering
    \includegraphics[width=\textwidth]{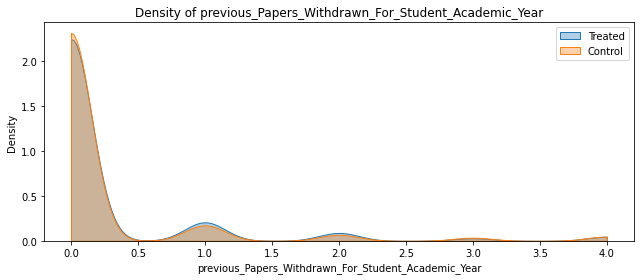}
    \caption{Previous courses withdrawn (last academic year).}
    \label{fig:papers_withdrawn}
  \end{subfigure}\hfill
  \begin{subfigure}[b]{0.48\textwidth}
    \centering
    \includegraphics[width=\textwidth]{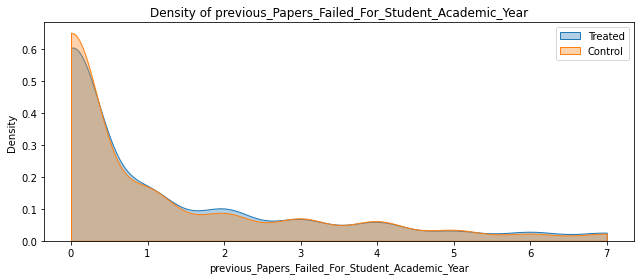}
    \caption{Previous courses failed (last academic year).}
    \label{fig:papers_failed}
  \end{subfigure}
  \caption{Density plots for selected baseline covariates (row 5).}
\end{figure}

\begin{figure}[htbp]
\centering
\begin{subfigure}{0.48\textwidth}
  \includegraphics[width=\linewidth]{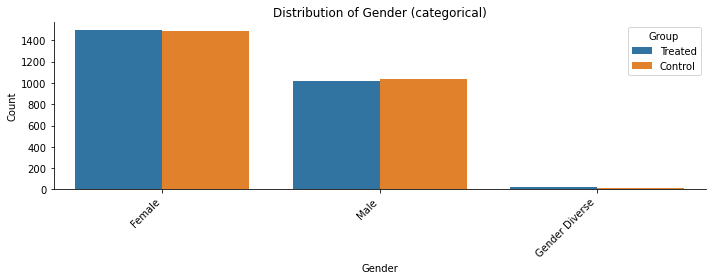}
  \caption{Gender}
      \label{fig:Gender}
\end{subfigure}
\begin{subfigure}{0.48\textwidth}
  \includegraphics[width=\linewidth]{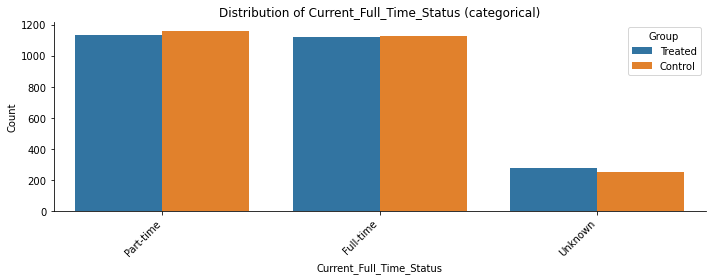}
  \caption{Full Time}
      \label{fig:fulltime}
\end{subfigure}

\begin{subfigure}{0.48\textwidth}
  \includegraphics[width=\linewidth]{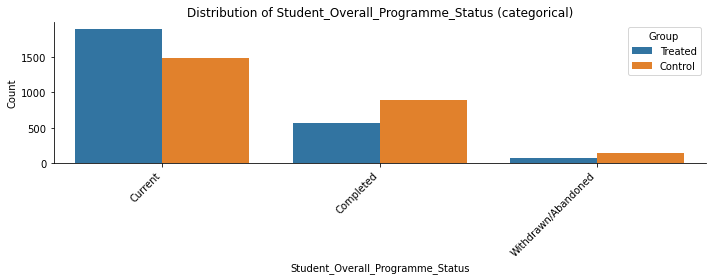}
  \caption{status}
\end{subfigure}  
\begin{subfigure}{0.48\textwidth}
  \includegraphics[width=\linewidth]{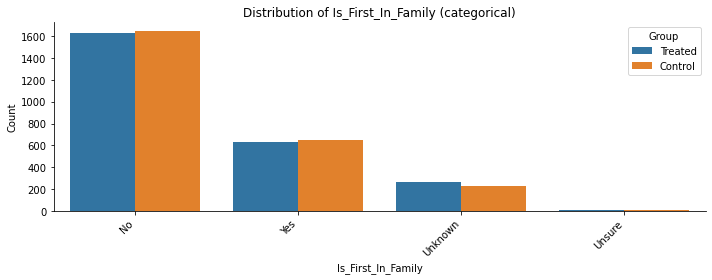}
  \caption{first in family}
  \label{fig:first}
\end{subfigure}

\begin{subfigure}{0.48\textwidth}
  \includegraphics[width=\linewidth]{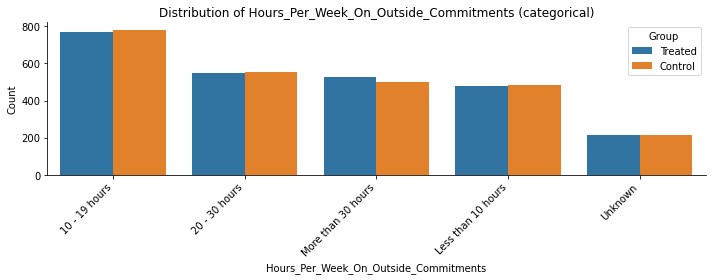}
  \caption{working hours}
  \label{fig:working}
\end{subfigure}
\begin{subfigure}{0.48\textwidth}
  \includegraphics[width=\linewidth]{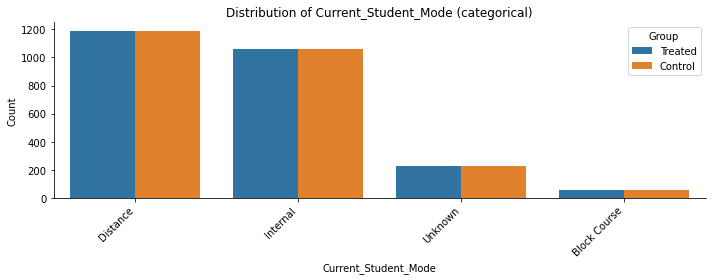}
  \caption{mode}
  \label{fig:mode}
\end{subfigure}

\caption{Summary of PSM categorical matching diagnostics.}
\end{figure}

\begin{figure}[htbp]
\centering

\begin{subfigure}{0.68\textwidth}
  \includegraphics[width=\linewidth]{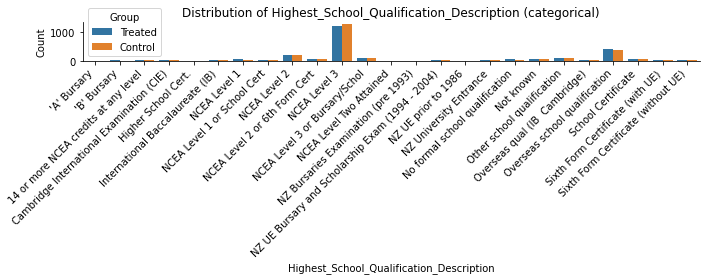}
  \caption{entry qualification}
  \label{fig:entryqualification}
\end{subfigure}

\begin{subfigure}{0.68\textwidth}
  \includegraphics[width=\linewidth]{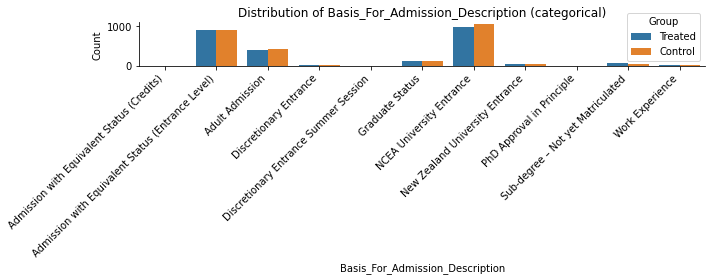}
  \caption{basis for admission}
  \label{fig:basis}
\end{subfigure}

\begin{subfigure}{0.68\textwidth}

  \includegraphics[width=\linewidth]{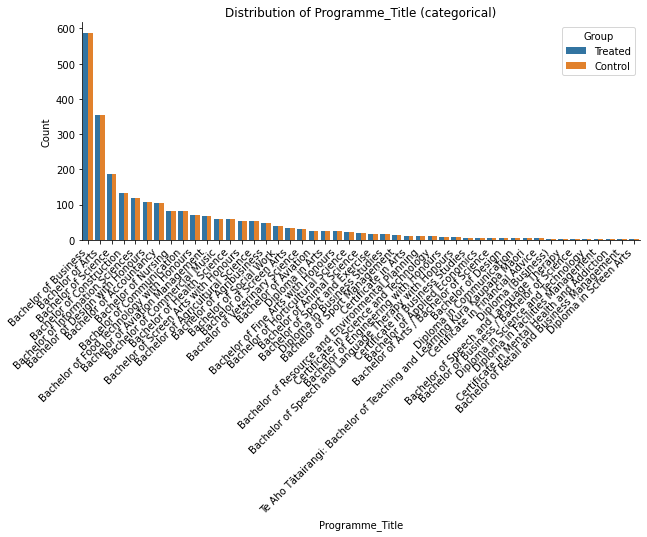}
  \caption{programme}
  \label{fig:programme}
\end{subfigure}
\caption{Summary of PSM categorical matching diagnostics.}
    \label{fig:categorical_diagnostics}
\end{figure}

Here, the diagnostic plots that assess the quality of the
propensity score matching procedure are presented (Figures \ref{fig:continuous_row1} -  \ref{fig:categorical_diagnostics}. The validity of the causal claims in
this study depends on treated and matched control students being comparable
on observed pre–treatment characteristics. 
The density plots display kernel-smoothed distributions for key continuous
covariates, including prior grades, credits and papers passed or failed,
age, and qualification progress. In each plot, the treated group and the
matched control group show substantial overlap, indicating that the matching
algorithm successfully aligned the distributions that were imbalanced in the
raw data. Bar plots for categorical covariates (for example gender, study
mode, first-in-family status, and basis for admission) further show that the
proportions in each category are very similar across groups. Taken together,
these diagnostics confirm that post–match covariate balance is acceptable
and that the estimated treatment effects are unlikely to be driven by
observed pre–existing differences.

\section{Rosenbaum sensitivity analysis}
\label{app:rosenbaum}
--------------------------------------------------------

Propensity score matching adjusts for observed covariates but cannot rule
out hidden confounding. To examine the robustness of the main findings to
unobserved factors, we computed Rosenbaum Wilcoxon signed–rank sensitivity
bounds for the outcomes with statistically significant doubly robust ATEs.
Table~\ref{tab:rosenbaum-summary} summarises the bounds across a grid of
values of the sensitivity parameter $\Gamma$. A value of $\Gamma = 1$
corresponds to no hidden bias, meaning that matched students have equal odds
of receiving support. Larger values represent the extent to which an
unobserved covariate could multiply those odds within pairs. The results
show that the reductions in course failures and the improvements in
cumulative grades and pass rates remain statistically significant for
moderate levels of hidden bias (up to $\Gamma \approx 2.5$ for some
outcomes). An unobserved factor would therefore need to more than double the
odds of treatment assignment to overturn the main conclusions.

\begin{table}[htbp]
\centering
\caption{Summary of Rosenbaum sensitivity analysis for selected outcomes with statistically significant values. $\Gamma$ is the sensitivity parameter; the last
column reports the largest value in the grid for which the worst–case
one–sided $p$–value remains below 0.05.}
\label{tab:rosenbaum-summary}
\begin{tabular}{lcc}
\hline
Outcome & $\Gamma$ grid & Max.\ $\Gamma$ with $p_{\text{upper}}<0.05$ \\
\hline
Courses failed (intervention year)              & 1.0, 1.25, 1.5, 2.0, 2.5 & 2.5 \\
Cumulative mean grade      & 1.0, 1.25, 1.5, 2.0, 2.5 & 2.5 \\
Cumulative deviation from class mean              & 1.0, 1.25, 1.5, 2.0, 2.5 & 2.5 \\
\hline
\end{tabular}
\end{table}

\end{document}